\documentclass[aps,pre,twocolumn,float]{revtex4}
\usepackage{amsmath, bm}
\usepackage{pstricks}
\usepackage{pst-node}
\usepackage[ansinew]{inputenc}
\usepackage{amssymb,amsmath}
\usepackage{amsfonts}
\usepackage{epsfig}
\usepackage[english]{babel}
\usepackage{graphics}
\usepackage{graphicx}
\usepackage{amsmath,bm,epsfig}
\let \nn  \nonumber
\usepackage{pstricks}
\usepackage{pst-node}
\usepackage[ansinew]{inputenc}
\usepackage{amssymb,amsmath}
\usepackage{multirow}

\def\a{\alpha}
\def\b{\beta}
\def\e{\varepsilon}

\def\o{\omega}

\def\be{\begin{equation}}       \def\ba{\begin{array}}

\def\ee{\end{equation}}         \def\ea{\end{array}}

\def\bea {\begin{eqnarray}}      \def\eea {\end{eqnarray}}

\def\bean{\begin{eqnarray*}}    \def\eean{\end{eqnarray*}}

\def\eps{\varepsilon}

\def\RA {\ \Rightarrow\ }

\newtheorem{exi}{Example}

\begin{document}

\title{Energy spectrum of ensemble of weakly nonlinear gravity-capillary waves on a fluid surface}
\author{Elena Tobisch}
 \email{Elena.Tobisch@jku.at}
  \affiliation{Institute for Analysis, Johannes Kepler University, Linz, Austria}

   \begin{abstract}
   
In this Letter we regard nonlinear gravity-capillary waves  with parameter of nonlinearity  being $\varepsilon \sim 0.1 \div 0.25$.
For this nonlinearity time scale separation does not occur and kinetic wave equation does not hold.
An energy cascade in this case is built at the dynamic time scale (D-cascade) and is computed by the
 increment chain equation method  first introduced in \emph{Kartashova, \emph{EPL} \textbf{97} (2012), 30004.} We compute for the
   first time an analytical expression for the energy spectrum   of nonlinear gravity-capillary waves
as an explicit function depending on the ratio of surface tension to the gravity acceleration. It is shown that its two limits - pure capillary and pure gravity  waves on a fluid surface - coincide with the previously obtained results.

We also discuss relations of the model of D-cascade with a few known models used in the theory of nonlinear waves such as Zakharov's equation, resonance of the modes with nonlinear Stokes corrected frequencies and Benjamin-Feir index.  These connections are crucial in the understanding and forecasting specifics of the energy transport in a variety of multi-component wave dynamics, from oceanography to optics, from plasma physics to acoustics.
\end{abstract}

\maketitle
\section{Introduction}

Until recently, the notion of "energy cascade" in a weakly nonlinear wave system was traditionally associated with \emph{kinetic} wave turbulence theory (WTT) where an energy spectrum is a stationary solution of the wave kinetic equation. The wave kinetic equation was first introduced in 1962 by Hasselmann, \cite{has1962}, and its first stationary solution (for capillary waves) was found  in 1967 by Zakharov and Filonenko, \cite{zak2}. Later on, their  method of finding stationary solutions was generalized to various weakly nonlinear wave systems with dispersion, \cite{ZLF92}.

Kinetic equation can be solved numerically for any nontrivial dispersion function $\o$ with given dispersion relation $\o=\o(\mathbf{k})$, where $\mathbf{k}$ is wave vector. In the particular case of dispersion function of the form $\o \sim k^{\a}, \, k=|\mathbf{k}|,$ with $\a>1$, kinetic WTT given analytical prediction for the energy spectrum in the form of power law $\sim k^{-\b}, \b>0,$ where $\b$ is different for wave systems with different dispersion function but does not depend on the excitation parameters.

The prediction hold at the so-called inertial interval where forcing and dissipation are balanced so that within this interval energy is conserving (it is assumed that the pumping and dissipation are spaced far apart in Fourier space). The basic physical mechanism leading to the formation of kinetic energy cascade (K-cascade), is $s$-wave resonant interactions of linear Fourier modes
\be
A(t/\varepsilon^{s-2}) \exp{i[kx-\o  t]}\ee
 with slowly changing amplitudes. The $s$-wave resonances occur independently at different time scales $t/\varepsilon^{s-2}, \, s\ge3,$ and $0<\varepsilon\ll 1$ a small parameter; e.g. $\varepsilon \sim 10^{-2}$ for water waves.

Rapid technological progress in the field of measurement methods and measuring techniques allowed the past two decades, systematic study of the spectrum in various fluid systems. The experimental data turned out to be rather contradictory, e.g.: energy spectrum is not formed - rather  energy exchange within a small set of Fourier modes; energy spectrum and power law are observed but the exponent differs from the predicted by kinetic WTT; the exponent depends on the parameters of initial excitation; inertial interval does not exist,  etc. Without claiming to be exhaustive we give a few references to the most  thorough and credible experiments of recent years, \cite{SWW-09,ABKL09,ParEx3,ShXP12,Erik}.  A very respectable list of references can be found in a recent review of Newell and Rumpf, \cite{NR11}.

Some of these effects have found their explanation in the frame of the \emph{discrete} WTT, \cite{K06-1,K09b}; for instance, the absence of the inertial interval is due to the non-locality of resonant interaction, for some types of dispersion functions. Locality of interaction in kinetic WTT is understood as following: only the interaction of waves with wave lengths of the same  order is allowed. However, is is know for more than 20 years \cite{PHD1,PRL}, that, say, capillary waves with wave lengths of order $k$ and $k^3$ can interact directly, i.e. build a joint resonance triad; more examples can be found in \cite{CUP}.

The  model of the energy spectra formation in the wave systems with weak and moderate nonlinearity
allowing to resolve the observed experimental differences over the shape of the energy spectrum with the predictions made for K-cascade was first proposed in 2012 by Kartashova,
\cite{K12a}. In this model the triggering physical mechanism for an energy cascade formation is modulation instability (MI), and corresponding energy cascade is called  dynamical cascade (D-cascade); a D-cascade which is a sequence of distinct modes in Fourier space. The use of the specially developed increment chain equation method (ICEM) allows to compute energy spectrum of a D-cascade.

Energy spectrum in the D-model is a solution of so-called chain equation. It connects frequencies and amplitudes of two adjacent modes in D-cascade. Energy spectra for capillary and surface water waves (with dispersion functions being $\o^2=\sigma k^3$ and $\o^2=g k$ consequently) are computed in \cite{K12a}, for different values of a small parameter $\varepsilon \sim 0.1 \div 0.4$ chosen as a ratio of the wave amplitude to the wave length.

In this Letter we give a brief sketch of the ICEM and compute energy spectrum of ensemble of weakly nonlinear gravity-capillary waves with dispersion function $\o^2=g k+\sigma k^3$. We demonstrate also the intrinsic mathematical connections between D-model and other models describing nonlinear wave interaction at the same time and space scales: Zakharov's equation, resonances of nonlinear Stokes waves and Benjamin-Feir index.

\section{Increment Chain Equation Method (ICME)}

 The physical mechanism underlying formation of a D-cascade is modulation instability which can be described as
the decay of a carrier wave  $\o_0$ into two side-bands $\o_1, \, \o_2$:
\bea
\o_1 + \o_2 = 2\o_0, \quad
\vec{k}_1+\vec{k}_2=2\vec{k}_0+ \Theta, \label{ModInst}\\
 \o_1=\o_0 + \Delta \o, \, \o_2=\o_0 - \Delta \o, \,  0<\Delta \o \ll 1. \label{Delta-Omega} \eea
A wave train with initial real amplitude $A$, wavenumber $k=|\vec{k}|$, and frequency $\o$  is modulationally unstable if
\be
0 \le {\Delta \o}/{A k\o} \le \sqrt{2}\label{interv-inst}.
\ee
Eq.(\ref{interv-inst}) describes an instability interval for the wave systems with a small nonlinearity of order of  $\e\sim 0.1$ to 0.2, first obtained in \cite{BF67}. It is also established for gravity surface waves that the most unstable modes in this interval satisfy the condition
\be \label{BFI-incr}
\Delta \o / A k \o =1.
\ee

The essence of the ICEM is the use of (\ref{BFI-incr}) for computing the frequencies of the cascading modes. At the first step of the D-cascade, a carrier mode has  frequency $\o_0$ and the distance to the next cascading mode $(\Delta \o)_1=|\o_0 - \o_1|$ with frequency $\o_0$ is chosen in such a way that  condition (\ref{BFI-incr}) is satisfied, i.e. $|\o_0 - \o_1|=A_0k_0\o_0$.
  At the second step of the D-cascade,
a carrier mode has the frequency $\o_1$, the distance to the next cascading mode  $(\Delta \o)_2=|\o_1 - \o_2|$  is chosen so that $|\o_1 - \o_2|=A_1k_1\o_1$, and so on.

In this way, a recursive relation
between neighboring cascading modes can easily be obtained:
\be \label{1}
\sqrt{p}_n A_n=  A(\o_n \pm \o_n A_n k_n)
\ee
Here notation $p_n$ is chosen for the fraction of  energy transported from the cascading mode $A_n=A(\o_n)$ to the cascading mode $A_{n+1}=A(\o_{n+1})$, i.e. $A_{n+1}=\sqrt{p_n}A_n$. This fraction $p$ is called cascade intensity, \cite{K12a}.

The Eq.(\ref{1}) describes two chain equations: one chain equation with "+" for direct D-cascade with $\o_n< \o_{n+1}$ and another chain equation  for inverse D-cascade with $\o_n> \o_{n+1}$.

Speaking generally, the cascade intensity $p_n=p_n(A_0, \o_0, n)$  might be a function of the excitation parameters $A_0, \o_0$ and the step $n$. However, as in a lot of experiments it is established that $p_n$ depends only on the excitation parameters and does not depend on the step $n$, all the formulae below are given for the case of the constant cascade intensity. Accordingly, notation $p$ is used instead of the notation  $p_n$. This means in particular that $A_{n+1}=\sqrt{p}A_n=p^{n/2}A_0$ and as energy $E_n \sim A_n^2$ it follows $E_n \sim p^n A_0^2$, i.e. energy spectrum of the amplitudes of the D-cascade has exponential form.

Taking Taylor expansion of the RHS of the chain equation and regarding only two first terms of the resulting infinite series, one can derive an ordinary differential equation describing stationary amplitudes of the cascading modes.
The
consequent steps of the ICEM are given below.

1) Relation between neighboring amplitudes :
\be \label{gen1}
A_{n+1}=\sqrt{p} A_n
\ee

2) Condition for maximal increment:
\be \label{gen2}
\frac{|\o_{n+1}- \o_n|}{f(\o_n A_n k_n)}= 1,
\ee
where $f(\o_n A_n k_n)$ is known  function of the product $\o_n A_n k_n$. For instance, $f(\o_n A_n k_n)=\o_n A_n k_n$ for gravity surface waves with small parameter of the order of $0.1\div 0.2$. Examples for bigger nonlinearity and also for other wave types can be found in
\cite{DY79,Hog85}.

3) Chain equations:
\bea \label{gen3}
A_{n+1}\equiv A(\o_{n+1})=\nn \\
A(\o_n\pm f(\o_n A_n k_n))  = \sqrt{p} A(\o_n)
\eea
where "+" should be taken for direct cascade and "-" for inverse cascade.

4) approximate ODE(s) on amplitude $A_n$: \be \label{gen4} \sqrt{p} A_n \approx A_n \pm A_n^{'}f(\o_n A_n k_n). \ee

5) Discrete energy spectrum: $ E_n \sim A_n^2$.

6) Spectral density: \be S(\o)=|\lim_{n\rightarrow \infty}\frac{ d \, E_n}{d \, \o_n}|.\ee

In particularly, the formula  below gives explicit expression for the computation of the wave amplitudes (for direct cascade) in the case of a small initial nonlinearity $\eps\sim 0.1 \div 0.25$:
\be
A(\o_n) = (\sqrt{p}-1) \int \frac{d\,  \o_n}{\o_n \cdot k(\o_n)} . \label{A-gen-omega}
\ee
Remark. It is known that if the autocorrelation function involves temporal measurements at a single point, the power spectrum has the units $m^2/Hz$. It is easy to check that spectral density $S(\o)$ has correct units.
Indeed,
as we compute the amplitudes at single points, amplitudes
 $A(\o_n)$ have units $m$, than their squares $A(\o_n)^2$ have units $m^2$ and spectral density $S(\o)$ has units $m^2/Hz$.

Let us illustrate the method described above with  an example  of gravity surface waves, with weak nonlinearity, $\e \sim 0.1\div 0.25$ first computed in  \cite{K12a}.

\emph{Example: Gravity surface waves.} In this case $\o^2=gk$, and  $f(\o_n A_n k_n)=\o_n A_n k_n$, \cite{BF67},  which yields
\bea
|(\Delta \o)_n|/\o_n A_n k_n =1.
 \RA  \\
\sqrt{p}-1 \approx \pm  A_n^{'}\o_n^3  \RA\\
A_n^{(Dir)}=g\, \frac{(1-\sqrt{p})}{2} \o_n^{-2}+C^{(Dir)},\\
A_n^{(Inv)}=-g\, \frac{(1-\sqrt{p})}{2} \o_n^{-2} +C^{(Inv)}, \\
E(\o_n)^{(Dir)} \sim \Big[ g\, \frac{(1-\sqrt{p})}{2} \o_n^{-2}+C^{(Dir)} \Big]^2, \\
E(\o_n)^{(Inv)} \sim \Big[ -g\, \frac{(1-\sqrt{p})}{2} \o_n^{-2} +C^{(Inv)} \Big]^2, \\
\mbox{with  } C^{(Dir)} = A_0 - g\,\frac{(1-\sqrt{p})}{2} \o_0^{-2}, \\
 C^{(Inv)} = A_0 + g\,\frac{(1-\sqrt{p})}{2} \o_0^{-2}.
\eea
As we assume that $p=$const, specific choice of the excitation  parameters allows us to compute explicitly the cascade intensity $p$
as the function of the excitation parameters $A_0, \, \o_0.$ Indeed, let us regard for instance the case of direct cascade and
the choice of excitation parameter is such that $C^{(Dir)}=0$. Then
\bea
A_0 - g\,\frac{(1-\sqrt{p})}{2} \o_0^{-2}=0 \RA \nonumber \\
p=[1-(2A_0\o_0^2)/g]^2.
\eea
Accordingly, $E(\o_n)^{(Dir)}=(A_0\o_0\o_n^{-2})^2 \sim \o_n^{-4}$ and spectral density reads $S(\o)_{grav} \sim \o^{-5}.$ This corresponds to the celebrated Phillips spectrum, \cite{Phil58}, and also to the real oceanic measurements  coined in the JONSWAP wave spectrum for wind-generated waves. The JONSWAP spectrum is the standard wave spectrum input used for in practical engineering, e.g. for practical fatigue calculation for off-shore structures,

Similar computation for capillary waves with small nonlinearity yield $E(\o_n)^{(Dir)}\sim \o_n^{-4/3}$ and $S(\o)_{cap} \sim \o^{-7/3}.$

\section{D-spectra of gravity-capillary waves}
\subsection{Computation of the spectrum}
In this case computation of $A_n=A(\o_n)$ is too tedious and is omitted here. Instead, we compute  $A_n=A(k_n)$ by changing integration variable in (\ref{A-gen-omega}); some preliminary  computations are necessary:
\bea
\o(k)=\sqrt{g \, k+\sigma k^3} \quad \RA \quad \o{'}_k = \\
\frac{g+3\sigma k^2}{2\sqrt{g \, k+\sigma k^3}} \RA \\
\frac{\o{'}_k}{\o(k)\cdot k}= \frac{g+3\sigma k^2}{(2\sqrt{g \, k+\sigma k^3})(\sqrt{g \, k+\sigma k^3})k}=\\
 \frac{g+3\sigma k^2}{2k(g \, k+\sigma k^3)} \RA \\
A_{gr-cap}(k)=(\sqrt{p}-1) \int{\frac{\o'_k \, d k}{\o(k) k}}=\\
\frac {(\sqrt{p}-1)}{2} \int{\frac{(g+3\sigma k^2) \, d k}{k(g \, k+\sigma k^3)}}.\label{integral-A-spect-grav-cap-k}
\eea
This indefinite integral can be computed explicitly:
\be
\int{\frac{(g+3\sigma k^2) \, d k}{k(g \, k+\sigma k^3)}}=\\
2\sqrt{\frac{\sigma}{g}} \arctan{(\sqrt{\frac{\sigma}{g}}k)} - k^{-1} + const.
\ee
which yields (for direct cascade)
\bea
A_{gr-cap}(k_n)=\nonumber \\
\frac {(1-\sqrt{p})}{2}\big[ k_n^{-1}- 2\sqrt{\frac{\sigma}{g}} \arctan{(\sqrt{\frac{\sigma}{g}}k_n)}  \big] \nonumber \\
- \frac {(1-\sqrt{p})}{2}\big[  k_0^{-1}- 2\sqrt{\frac{\sigma}{g}} \arctan{(\sqrt{\frac{\sigma}{g}}k_0)} \big] + A_0.
\eea
Keeping in mind that cascade intensity $p$ is constant it follows that
\bea
&& A_{gr-cap}(k_n) \sim \big[ k_n^{-1}- 2\sqrt{a} \arctan{(\sqrt{a}k_n)}  \big], \label{A-spect-grav-cap-k}\\
&& E_{gr-cap}(k_n) \sim \big[ k_n^{-1}- 2\sqrt{a} \arctan{(\sqrt{a}k_n)}  \big]^2,\label{E-spect-grav-cap-k} \\
&& S(k)_{gr-cap}\nonumber \\
&&\sim |\left(\frac{1}{k^2}+\frac{2 a^2}{1+a^2 k^2}\right) \left(\frac{1}{k}-2 a \arctan(a\, k)\right)|,\label{S-spect-grav-cap-k}
\eea
where notation $a=\frac{\sigma}{g}$ is used.

The D-cascade among gravity-capillary water waves with $\varepsilon \sim 0.1 $ is formed at the time scale of order of dozen of seconds, \cite{K13a}; for instance, for a wave with wavelength 10 cm corresponding time scale is 25 seconds, and D-cascade would be easy to observe in a laboratory experiment.

\subsection{Consistency check}
Energy spectra for pure gravity and pure capillary waves in \cite{K12a} have been obtained in the form $A=A(\o)$.
To check consistency of (\ref{A-spect-grav-cap-k}) with the results obtained above for gravity-capillary waves   we have  to rewrite them in the form $A=A(k)$ as following:
\be
A(k) =  (\sqrt{p}-1) \int \frac{\o{'}_k d k}{\o(k)\cdot k} \label{A-gen-k}.
\ee
This yields
\emph{for surface gravity waves}
\bea
  A_{grav}(k)=(\sqrt{p}-1) \int \frac{\o{'}_k d k}{\o(k)\cdot k} =\nonumber\\
   (\sqrt{p}-1)\int \big(\frac{\sqrt{g}}{2\, k^{1/2}} \frac{1}{ \sqrt{g} \, k^{1/2} k}\big)d k =\nonumber\\
   \frac{(\sqrt{p}-1)}{2} \int \frac{d k}{k^2} \RA \nonumber\\
A_{grav}(k)=  \frac{(1-\sqrt{p})}{2}(k^{-1}-k_0^{-1}) + A_0 \label{A-spect-grav-k}
\eea
and \emph{for capillary waves}
\bea
A_{cap}(k)=(\sqrt{p}-1) \int \frac{\o{'}_k d k}{\o(k)\cdot k} =\nonumber\\
(\sqrt{p}-1) \int \big(\frac{3 \,\sqrt{\sigma}\, k^{1/2}}{2} \frac{1}{\sqrt{\sigma} k^{3/2} \, k} \big)d k= \nonumber\\
\frac{3(\sqrt{p}-1)}{2} \int \frac{d k}{k^2} \RA \nonumber\\
A_{cap}(k) = \frac{3(1- \sqrt{p})}{2} (k^{-1}-k_0^{-1})+ A_0 \label{A-spect-cap-k}
\eea
In order to avoid tedious calculations, let us regard a special choice of excitation parameters such that $A(k) \sim k^{-1}$, then
for this case we get
\bea
E_{grav}(k) \sim k^{-2}, \quad S_{grav}(k) \sim k^{-3}  \label{*}\\
E_{cap}(k) \sim k^{-2}, \quad S_{cap}(k) \sim k^{-3}\label{**}
\eea
Let us now regard the expression under the integral in (\ref{integral-A-spect-grav-cap-k}) for two limiting cases: (a) $\sigma \rightarrow 0$, and (b)  $g  \rightarrow  0$. In the first case,
 the integral (\ref{integral-A-spect-grav-cap-k}) transforms into
$ \int{k^{-2} d k}$ and consequently
\be
A_{gr-cap}(k)= \frac {(1-\sqrt{p})}{2}\big[ k^{-1}  - k_0^{-1}\big]+ A_0 =   A_{grav}(k).
\ee
In the second case,   the integral (\ref{integral-A-spect-grav-cap-k}) transforms into
$ \int{3k^{-2} d k}$ and consequently
\be
A_{gr-cap}(k)= \frac {3(1-\sqrt{p})}{2}\big[ k^{-1}  - k_0^{-1}\big]+ A_0 =   A_{cap}(k)
\ee
This means the expression for the energy spectrum of gravity-capillary waves is consistent with previously obtained results for pure gravity and pure capillary waves, i.e. the D-model itself is consistent.

Another important check follows from the standard relation $\int S(\o)d\,\o=\int S(k)d\,k$ (in one space dimension). Rewriting it as
 $S(k)=S(\o)\frac{d \, \o}{d \, k}$  we can compute $S(k)_{grav}$ and $S(k)_{cap}$:
\be
S(k)_{grav} \sim k^{-1/2} k^{-5/2}=k^{-3} \ee
and
\be
S(k)_{cap} \sim k^{-7/2} k^{1/2}=k^{-3}  \ee
 which is
in accordance with formulae (\ref{*}),(\ref{**}).

\section{Connection of D-model with other models}
On different scales in time and space there are many models describing various phenomena and processes in nonlinear wave interaction. Some of these models have the same time and space scale as D-cascade. In this section we
show that direct  mathematical relationship  between D-model and a few other known models exists.

(I) The computation of the D-cascade spectra demonstrated above and in \cite{K12b} has been performed in the frame of nonlinear Schr\"{o}dinger equation or its modifications. As modulation instability exists in other evolutionary dispersive nonlinear partial differential equations, e.g. in  generalized versions of Korteweg-de Vries equation \cite{BJK11,MatJohn13}, Hasegawa-MIma equation \cite{AndSmol} and others, the ICEM can be directly applied also for these equations. All the difference between different equation would be "hidden" in the form of the chain equation.

 (II) The  computation kindly provided to us by Miguel Onorato in the general discussion at the Workshop "Wave Turbulence" (Ecole de Physique, Les Houches, France, 2012)
 shows connection between D-cascade and Zakharov's equation.

Let us first rewrite (\ref{gen2}) as
\bea
\tilde \o_{R}=\o_0 +  \o_0 A_0 k_0, \label{inc-n1-dir-1}\\
\tilde  \o_{L}=\o_0 -  \o_0 A_0 k_0, \label{inc-n1-inv-1}
\eea
and  consider a system of 3 discrete waves using the following decomposition:
\be \label{copmos}
a_k=b_0\delta_k^0+b_L\delta_{k}^{L}+b_R\delta_{k}^{R}
\ee
Here $b_0$ is the carrier wave and $b_L$ and $b_R$
are the left and right sidebands, respectively; $L=k_0+\Delta k$
and $R=k_0-\Delta k$.

Assuming  $b_L$ and $b_R$
small with respect to the $b_0$ and  neglecting  nonlinear terms in
the sidebands amplitude, after substituting (\ref{copmos}) into Zakharov's equation
\be \label{zakh}
\frac{\partial a_1}{\partial t}+i \omega_1 a_1=
-i\int dk_{2,3,4} T_{1,2,3,4}a_2^* a_3 a_4 \delta_{12}^{34}
\ee
we get:
\be
\begin{split}
&\frac{db_0}{dt}+i\omega_0b_0=-i T_{0,0,0,0} |b_0|^2 b_0 \\
&\frac{db_L}{dt}+i\omega_Lb_L=-i 2T_{L,0,L,0} |b_0|^2 b_L-
i T_{L,R,0,0}b_0^2 b_R^* \\
&\frac{db_R}{dt}+i\omega_R b_R=-i 2T_{R,0,R,0} |b_0|^2 b_R-
i T_{R,L,0,0}b_0^2 b_L^*
\end{split}
\ee

If one is interested only in the interaction of each
sideband with the carrier wave, independently from the other, then one can
easily find the dispersion relation for $b_L$ and $b_R$.
Solution of the first equation is straightforward:
\begin{equation}
b_0=\tilde b_0 \exp[-i (\omega_0+T_{0,0,0,0} |\tilde b_0|^2)t]
\end{equation}
so that
\begin{equation}
\tilde \omega_0=\omega_0+T_{0,0,0,0} |\tilde b_0|^2
\end{equation}
$T_{0,0,0,0}=k^3$ but one should recall that the
Zakharov's equation is written for the wave action variable
which is related to the surface elevation $\eta_0$ as follows:
$\eta_0=\sqrt{2 k_0/\omega_0} b_0$, therefore the
dispersion relation is:
\be
\tilde \omega_0=\omega_0\left(1+\frac{1}{2}  k_0^2\eta_0^2\right)
\ee
and, neglecting the interaction between the two sidebands, we get:
\be
\begin{split}
&\frac{db_L}{dt}+i\omega_Lb_L=-i 2T_{L,0,0,L} |b_0|^2 b_L \\
&\frac{db_R}{dt}+i\omega_R b_R=-i 2T_{R,0,0,R} |b_0|^2 b_R
\end{split}
\ee
The nonlinear dispersion relation for the sidebands is:
\begin{equation}
\begin{split}
\tilde \omega_L=\omega_L+2 T_{L,0,0,L} |\tilde b_0|^2\\
\tilde \omega_R=\omega_R+2 T_{R,0,0,R} |\tilde b_0|^2
\end{split}
\end{equation}
The diagonal part of the coupling coefficient has the following form:
$T_{1,2,1,2}=k_1 k_2 min(k_1,k_2)$, therefore:
\begin{equation}
\begin{split}
T_{L,0,0,L}=k_0 k_L^2=k_0 (k_0-\Delta k)^2\\
T_{R,0,0,R}=k_0^2 k_R=k_0^2 (k_0+\Delta k)
\end{split}
\end{equation}
Therefore we get:
\begin{equation}
\begin{split}
\tilde \omega_L=\sqrt{(g (k_0-\Delta k)}+  \omega_0 (k_0-\Delta k)^2\eta_0^2\\
\tilde \omega_R=\sqrt{(g (k_0+\Delta k)}+\omega_0 k_0 (k_0+\Delta k) \eta_0^2
\end{split}
\end{equation}
Taylor expanding in $\Delta k$ and
neglecting nonlinear dispersive terms of cubic order, the resonant condition
becomes:
\bea
2  \tilde \omega_0-\tilde \omega_L-\tilde \omega_R=
 -\eta_0^2 k_0^2 \omega_0+\frac{\omega_0 \Delta k^2}{4 k_0^2}=0, \\
\RA \frac{\Delta k}{k_0}=2 k_0 \eta_0 = 2 \e, \label{steep1}
\eea
where $\e$ is the steepness of the carrier wave. As $2  {\Delta k}/{k_0}= {\Delta \omega}/{\omega_0}$, (\ref{steep1}) can be rewritten as
\begin{equation}
\frac{\Delta \omega}{\omega_0}= \e. \label{steep2}
\end{equation}
Eq.(\ref{steep2}) coincides with (\ref{BFI-incr}) and therefore corresponds
to the maximum of instability in the Benjamin-Feir
instability curve.

Thus we have shown that each step of the D-cascade, (\ref{gen3}), describes an exact 4-wave resonance in the Zakharov equation among the modes with nonlinear Stokes corrected frequencies.

(III) The D-cascade as a whole can be regarded as a resonance cluster formed by a few connected exact resonances of nonlinear Stokes modes. This means that in order to deduce the chain equation we do not need the modulation instability; in fact, the MI is just a suitable mathematical language for describing an energy cascade in the focusing evolutionary NPDEs. The general mathematical object, describing energy cascades both for focusing and non-focusing NPDEs, is a resonance cluster formed by a few connected exact resonances of nonlinear Stokes modes.

Objects of this type are studied by the homotopy analysis method (HAM), \cite{Liao12}.  Recent application of the HAM allowed to describe  steady-state resonance of multiple wave interactions in deep water, \cite{LL14}; numerical simulation with Zakharov's equations demonstrate qualitative agreement with the results obtained by the homotopy analysis method.

In contrast to  perturbation methods usually applied for studying nonlinear problems, this method does not introduce a small parameter and works in realistic physical set-ups.

(IV) Freak or rogue waves is a quite popular subject in the last few decades when different models for describing their appearance have been suggested.
Three main types of models are used for describing  rogue
waves formation: linear (spatial focusing or focusing due to dispersion), weakly nonlinear (focusing due to modulation instability) and essentially
nonlinear wave interaction, \cite{SDP11}. In the real numerical models for weather and ocean wave field prediction, the so-called Benjamin-Feir index (\emph{BFI}) is successfully used for characterizing the probability of the freak wave appearance.

The \emph{BFI} is defined as the ratio of the wave steepness $k_0 A$ to the spectrum width $\Delta \o / \o_0$ and the probability of high waves occurrence is non-zero if $BFI = 1$ or bigger, i.e. beginning with
\be \label{BFI}
BFI= \frac{k_0 A}{\Delta \o / \o_0} = 1.
\ee
However, quite recently it was shown that  freak waves can also occur in the systems where the MI is absent and \emph{BFI}=0, \cite{BCDL2013}.

This  apparent contradiction is easy to explain if we note that (\ref{BFI}) is equivalent to the condition for the maximal increment
(\ref{gen2}) and consequently to the chain equation (\ref{gen3}) which in turn can be described without making use of modulation instability, as it was demonstrated in  (III).

\section{Discussion}
In this Letter we have demonstrated how to apply increment chain equation method for computing energy spectra of ensemble of weakly nonlinear gravity-capillary waves with dispersion function $\o^2=g k+\sigma k^3$ and small parameter $\varepsilon \sim 0.1 \div 0.25$.  The energy spectrum is computed analytically as a function of $g/\sigma$, see (\ref{S-spect-grav-cap-k}); D-spectra in the two limiting cases - pure gravity case, $\o^2=g\,k$, and pure capillary case, $\o^2=g\,k^3,$ coincide with the known results first presented in \cite{K12a}.

The D-cascade among gravity-capillary water waves is formed at the time scale of order of dozen of seconds and  can easily be observed in a laboratory experiment. Various characteristic  of a  D-cascade in this case - its direction, possible scenarios of cascade termination, etc. - can be studied analytically, similar to the case of pure gravity waves presented in \cite{K12b}. This is the work in progress.

We have also demonstrated that D-cascade, though being a novel model, is directly connected with other important topics widely studied in fluid mechanics, e.g.
resonance clustering of the modes with nonlinear Stokes corrected frequencies or criterion for the freak wave appearance.
This makes it possible to transfer ideas, concepts and approaches from one scientific area to another  and to study them in a new setting. The understanding of the connections between different models are crucial in the understanding and forecasting specifics of the energy transport in a variety of multi-component nonlinear wave systems appearing everywhere from oceanography to optics, from plasma physics to acoustics.

{\textbf{Acknowledgements.}} The author is very much obliged to  W.E. Farrell, Walter Munk and Miguel Onorato  for  useful comments and suggestions.
 This research has been supported by
the Austrian Science Foundation (FWF) under project
P22943.

\end{document}